\documentclass[a4,useAMS,usenatbib,usegraphicx]{mn2e}

\def\kms{\,km\,s$^{-1}$}
\def\aaps{A\&AS}
\def\apj{ApJ}
\def\apjl{ApJL}
\def\mnras{MNRAS}
\def\pasp{PASP}
\def\aj{AJ}
\def\araa{ARA\&A}
\def\aap{A\&A}
\def\apjs{ApJS}
\def\nat{Nature}
\def\pasj{PASJ}

  \title[Coma UV LF and SFR]{The UV luminosity function and star formation rate of the Coma cluster.}

  \author[L. Cortese et al.]{L. Cortese$^1$, G. Gavazzi$^2$, A. Boselli$^3$\\
  $^1$ School of Physics and Astronomy, Cardiff University, Cardiff CF24 3AA, UK. \\
  $^2$ Universit\'{a} degli Studi di Milano-Bicocca, Piazza della Scienza 3, 20126 Milano, Italy\\ 
  $^3$ Laboratoire d'Astrophysique de Marseille, UMR 6110 CNRS, 38 rue F.Joliot-Curie, F-13388  Marseille, France \\
    }

\begin{document}
\date{Accepted 2008 August 13.  Received 2008 August 13; in original form 2008 June 2}

\maketitle

\label{firstpage}

\begin{abstract}
We present estimates of the GALEX near-ultraviolet (NUV) and far-ultraviolet (FUV) luminosity functions 
(LFs) of the Coma cluster, over a total area of $\sim$9 deg$^{2}$ ($\sim$25 Mpc$^2$), i.e. from the cluster center to the virial radius. 
Our analysis represents the widest and deepest UV investigation of  a nearby cluster of galaxies made to date.
The Coma UV LFs show a faint-end slope steeper than the one observed in the local field. This difference, more evident in NUV,  is entirely due to the contribution of massive quiescent systems (e.g. ellipticals, lenticulars and passive spirals), more frequent 
in high density environments.
On the contrary, the shape of the UV LFs for Coma star-forming galaxies does not appear to be significantly different from that of the field, consistently with previous studies of local and high redshift clusters. 
We demonstrate that such similarity is only a selection effect, not providing any information on 
the role of the environment on the star formation history of cluster galaxies.
By integrating the UV LFs for star-forming galaxies (corrected for the first time for internal dust 
attenuation), we show that the specific star formation rate (SSFR) of Coma is significantly lower than the integrated SSFR of the field and that Coma-like clusters contribute only $<$7\% of the total SFR density of the local universe.
Approximately 2/3 of the whole star-formation in Coma is occurring in galaxies with M$_{star}<$10$^{10}$ M$_{\odot}$.
The vast majority of star-forming galaxies has likely just started its first dive into the cluster 
core and has not yet been affected by the cluster environment. The total stellar mass 
accretion rate of Coma is $\sim$(0.6-1.8) $\times$ 10$^{12}$ M$_{\odot}$ Gyr$^{-1}$, suggesting that 
a significant fraction of the population of lenticular and passive spirals observed today in Coma could originate from infalling galaxies accreted between $z\sim$1 and $z\sim$0. 
\end{abstract}

\begin{keywords}
galaxies: clusters:individual: Coma (Abell1656)--galaxies: fundamental parameters--galaxies: evolution--galaxies: luminosity function--ultraviolet: galaxies
\end{keywords}

\section{Introduction}
Star formation is probably the most fundamental of all astrophysical processes regulating galaxy evolution: not only the 
properties of galaxies depend on how their stars were formed, but the properties of 
the interstellar medium (ISM) are controlled to a large extent by the 
various feedback effects of star formation. 
The overall evolution of galaxies depends on the rate at which their interstellar gas is converted 
into stars and the star formation rate (SFR) depends on the rate at which diffuse interstellar matter 
is collected into star forming regions. Consequently, it is of great importance for galaxy evolution 
studies to understand how stars form and what determines their properties. 

The cluster environment is well known to affect both the star formation activity and morphology
of the member galaxies, resulting in the morphology-density (e.g. \citealp{dressler80,whitmore}) and star formation-density (e.g. \citealp{lewis02,gomez03,haanna,review}) relations.
The scale over which the number density of elliptical galaxies increases toward the center of rich clusters 
is very small ($\leq$ 0.5 Mpc), while the one over which the star formation activity and the 
fraction of late-type galaxies decrease is much 
larger (a few Mpc). This indicates that separate processes contribute 
to shaping these relations. The mechanism involved in the early-type central enhancement is probably 
connected to the early phases of structure formation (e.g. \citealp{delucia06,ellis97,stanford98,renzini06}), 
while the progressive quenching of the star formation and lack of spirals toward the center of rich clusters 
could represent a separate, diluted in space and time, process probably 
related to the cluster accretion history (e.g. \citealp{dress04car,review,big,poggianti06}).

Although the effects of the environment on the star formation activity of cluster galaxies are gradually being unveiled (e.g. \citealp{review,n4569,dress04car,harrassment,poggia99,poggianti06,vollmer01}), we still 
know very little about the contribution of clusters of galaxies to the whole star formation history (SFH) and the stellar mass growth of the universe. 
The cosmic SFR dramatically decreases from z$\sim$1 to the present epoch \citep{lilly96,madau96}, but we do not know if this drop varies with the environment \citep{kodama01_cl,cooper08,finn08}, what fraction of the cosmic SFR takes place in cluster galaxies and how this number evolves with redshift. 

In addition, a significant increase from $z\sim$1 to $z=$0 in the stellar mass density of the universe has 
been observed (e.g. \citealp{borch06,drory05,fontana06}). Interestingly very little growth is found in the stellar mass of blue sequence galaxies, and the 
vast majority of stars formed since $z\sim$1 appears to inhabit galaxies which, at the present epoch, lie in the red sequence (e.g. \citealp{cimatti06,bell07,brown07,faber07}).
This automatically implies that a significant fraction of star-forming galaxies has migrated from the blue to the red sequence. 
The mechanism responsible for the quenching of the star formation in blue galaxies is still unknown and remains one of the greatest challenges for modern extragalactic astronomy. However, it is interesting to note that a similar migration, from star-forming to passive galaxies, 
is usually observed in high density environments (e.g. \citealp{dEale}), perhaps suggesting that the environment might play an important role. 

In order to clarify these issues, it is therefore crucial to quantify the contribution of galaxy clusters to the total SFR and stellar mass budget of the universe, and their evolution with redshift. 
To reach this goal we need to determine the SFR, luminosity and mass functions of the cluster population up to the virial radius, avoiding bias in our determination of the global cluster properties by observing a population that is not representative of the whole cluster environment.
This kind of study has become possible only recently thanks to the advent of wide field surveys. 
In particular, the Galaxy Evolution Explorer (GALEX, \citealp{martin05}) is providing for the first time precise UV photometry of galaxies over large stretches of the sky, opening a new era of extragalactic UV astronomy.
The UV emission is dominated by young stars of intermediate masses (2$<M<$5 M$_{\odot}$, e.g. \citealp{kenn98}) and thus represents an ideal tracer to identify and quantify the star formation activity of clusters. 
Surprisingly, very few statistical studies of galaxy clusters in UV have been carried out so far and great part of what we know 
about UV properties of cluster galaxies is still based on pioneering observations by the FAUST telescope (e.g. \citealp{deharveng94,brosh97})
and the balloon-borne UV telescopes SCAP and FOCA (e.g. \citealp{andreon,COGA03,donas87,donas95}).
For example, we still lack an accurate 
determination of the UV LFs in high density environments. 
The FOCA UV LFs were in fact affected by large statistical errors due to the insufficient redshift coverage for UV-selected galaxies and to the uncertainty in the UV background counts.
Moreover, the only GALEX UV LF for a nearby cluster presented so far (Abell1367, \citealp{CORBGA05}) does not sample the whole cluster region and suffers from low number statistics at high luminosities. 
Thus, in order to investigate the UV properties of nearby clusters we carried out a wide panoramic survey in near-ultraviolet (NUV) and far-ultraviolet (FUV) of the nearby Coma cluster of galaxies. \\
Coma (Abell 1656) is one of the nearest richest clusters, and it is considered 
as the prototype of evolved, relaxed cluster of galaxies. 
In the last decades, it has been one of the main laboratory for environmental studies at all 
wavelengths (e.g. see \citealp{review} and references therein) 
and more recently it has become the focus of a Hubble/ACS Treasury survey 
\citep{carter08} making it an ideal target for a deep census of UV properties in a nearby cluster up to 
its virial radius.\\
In this paper, we present the determination of the near-ultraviolet (NUV) and far-ultraviolet (FUV) LFs, the total SFR of the Coma cluster and its contribution to the SFR density in the local universe.

We assume a distance modulus of $m-M$ = 35.0 mag for the Coma cluster, corresponding 
to a distance of 100 Mpc  and a scale of  1.67 Mpc deg$^{-1}$ for $H_{0}$=70 Mpc$^{-1}$ km s$^{-1}$.

\section{The Data}
GALEX provides FUV ($\lambda$ = 1539 \AA, $\Delta \lambda$ = 442 \AA) and NUV ($\lambda$ = 2316 \AA, $\Delta \lambda$ = 1069 \AA) images with a circular field of view of $\sim$0.6 deg radius. The spatial resolution is $\sim$4-5 arcsec. Details of the GALEX instrument can be found in \cite{martin05} and \cite{morrisey05,morrissey07}. 

\begin{figure}
\centering
\includegraphics[width=14cm]{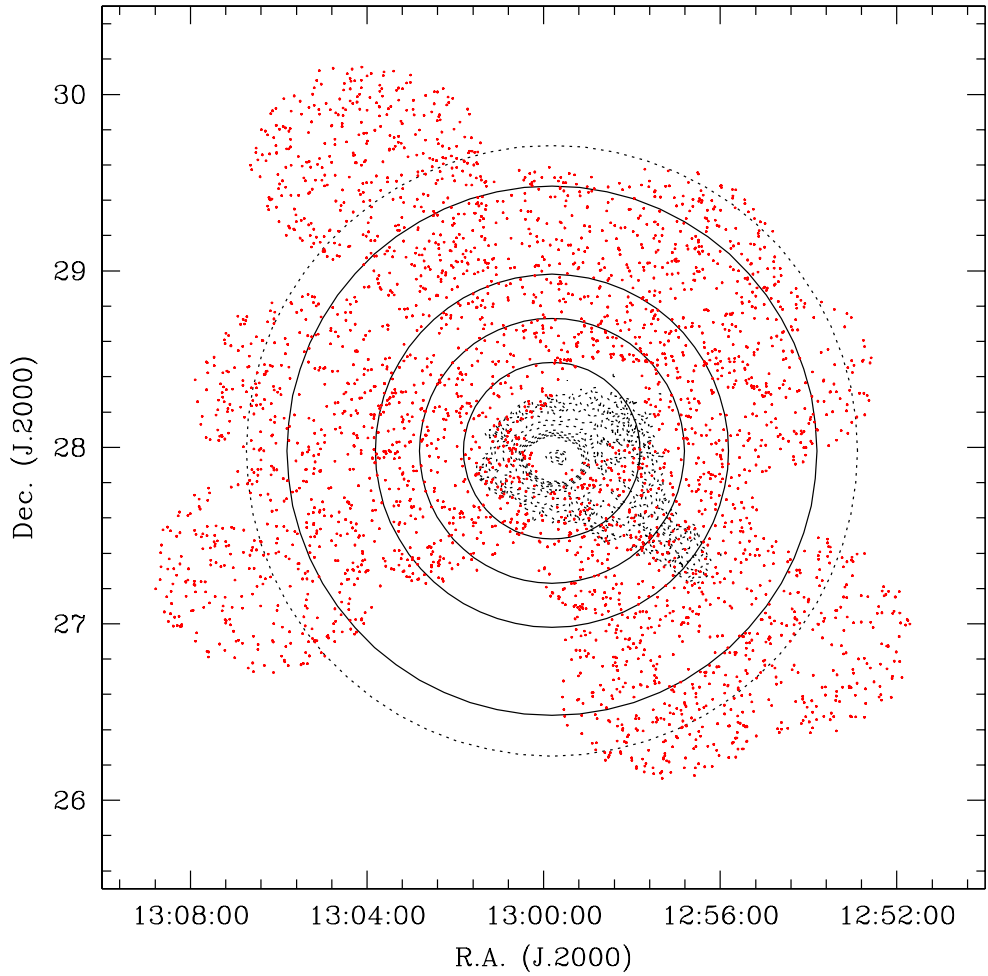}
\caption{\label{map}The Coma cluster region analyzed in this work. Red dots show all the NUV sources with $m_{AB}(NUV)<$21.5. Black 
contours indicate the Coma X-ray emission as observed by XMM and  circles indicate the apertures used to determine the LF in this work. The dotted circle is at the 
virial radius of Coma $\sim$2.9 Mpc, as determined by Lokas \& Mamon (2003).}
\end{figure}

The data analyzed in this paper are part of  a Cycle 1 Guest Investigator proposal (P.I. G. Gavazzi, proposal number: GI1-039). Twelve pointings have been originally approved but only 9 have been completed so far. In addition, in order to increase the GALEX coverage of the Coma cluster, we include in our analysis two more fields publicly available as part of the Nearby Galaxy Survey: NGC\_DDO154 and Coma\_SPECA. 
The coordinates and exposure times of the 11 fields are shown in Table \ref{fields}.  Their typical exposure time is $\sim$1500 sec in FUV and it varies between 1500 and 3000 sec in NUV. 
The total area covered  is $\sim$9 deg$^{2}$ (see Fig.~\ref{map}).
The major limitation of these observations  is that they do not include the center of Coma. The presence of an UV bright star near the cluster center makes this region unobservable in NUV. FUV observations of this region are present in the GALEX 
schedule but they have not yet been executed. The total area not observed near the center of Coma is $\sim$0.26 deg$^{2}$ (see Fig.~\ref{map}).

Each field has been reduced using the standard GALEX Data Analysis Pipeline \citep{morrissey07}.
Sources were detected and measured using SExtractor \citep{sex}. As the NUV images are significantly deeper than the FUV images, sources were selected, and their parameters determined, in the NUV. FUV parameters were extracted within the same apertures. 
To avoid artifacts present at the edge of the field, we considered only the central 0.58 deg radius from the field center. 
We used a variable SExtractor deblending parameter contrary to the standard GALEX pipeline, providing reliable magnitudes (MAGAUTO) also for very extended sources \citep{CORBGA05}. 
By comparing the flux estimates for objects detected in more than one field we find an 
average uncertainty in the NUV and FUV magnitudes of $\sim$0.15 mag. This value decreases to 
$\sim$0.07 magnitudes for objects brighter than $m_{AB}$(NUV)$\sim$19 mag.
Magnitudes are corrected for Galactic extinction using the \cite{schlegel98} reddening map 
 and the Galactic extinction curve of \cite{cardelli89}. 
The applied extinction corrections are 0.09 and 0.08 mag for the NUV and FUV bands, respectively. 
The 100\% completeness limit for a typical exposure time of 1500 sec is $m_{AB}\sim$  21.5 in both FUV and NUV \citep{xu}. However, in this work we will consider only objects brighter than 21 mag, because the poor redshift completeness makes impossible to investigate the properties of the 
LF at fainter luminosities.

The GALEX resolution is not sufficient to accurately separate stars and galaxies.
To address this issue we matched the GALEX catalog against the 
SDSS-DR6 \citep{sdss_DR6} observations of the Coma cluster. 
Stellar objects and artifacts in the SDSS plates are sometimes erroneously present in both star and galaxy SDSS catalogue.
For this reason, firstly we cross-matched our sample with the SDSS stellar catalogue using a search radius of 6 arcsec; secondly we cross-correlated the remaining UV detections with the SDSS galaxy catalogue, using the same impact parameter. 
Finally we cross-matched our catalogue with NED in order to obtain additional redshifts for the UV selected sample. 
A total number of 762 galaxies in FUV and of 1640 galaxies in NUV with $m_{AB}<$21 mag have been detected in the $\sim$9 deg$^{2}$ analyzed in this work.

\section{The NUV and FUV luminosity functions}
\begin{figure}
\centering
\includegraphics[width=14cm]{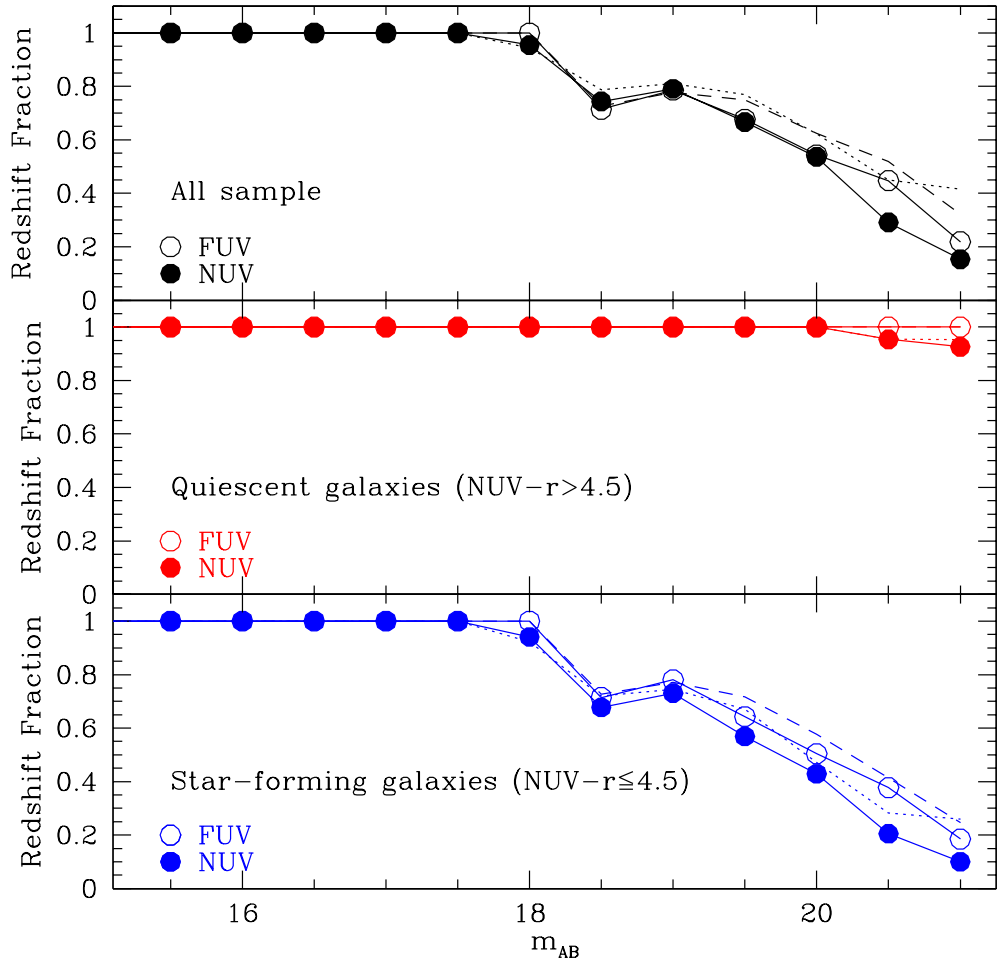}
\caption{\label{completeness}The NUV (filled circles) and FUV (empty circles) redshift completeness for the whole sample (upper panel), for quiescent (middle panel) and star forming (bottom panel) systems.
The dotted and dashed lines show the redshift completeness, in NUV and FUV respectively, after 
the exclusion of background galaxies selected using the colour-colour plot in Fig. 3.}
\end{figure}

\begin{figure}
\centering
\includegraphics[width=14cm]{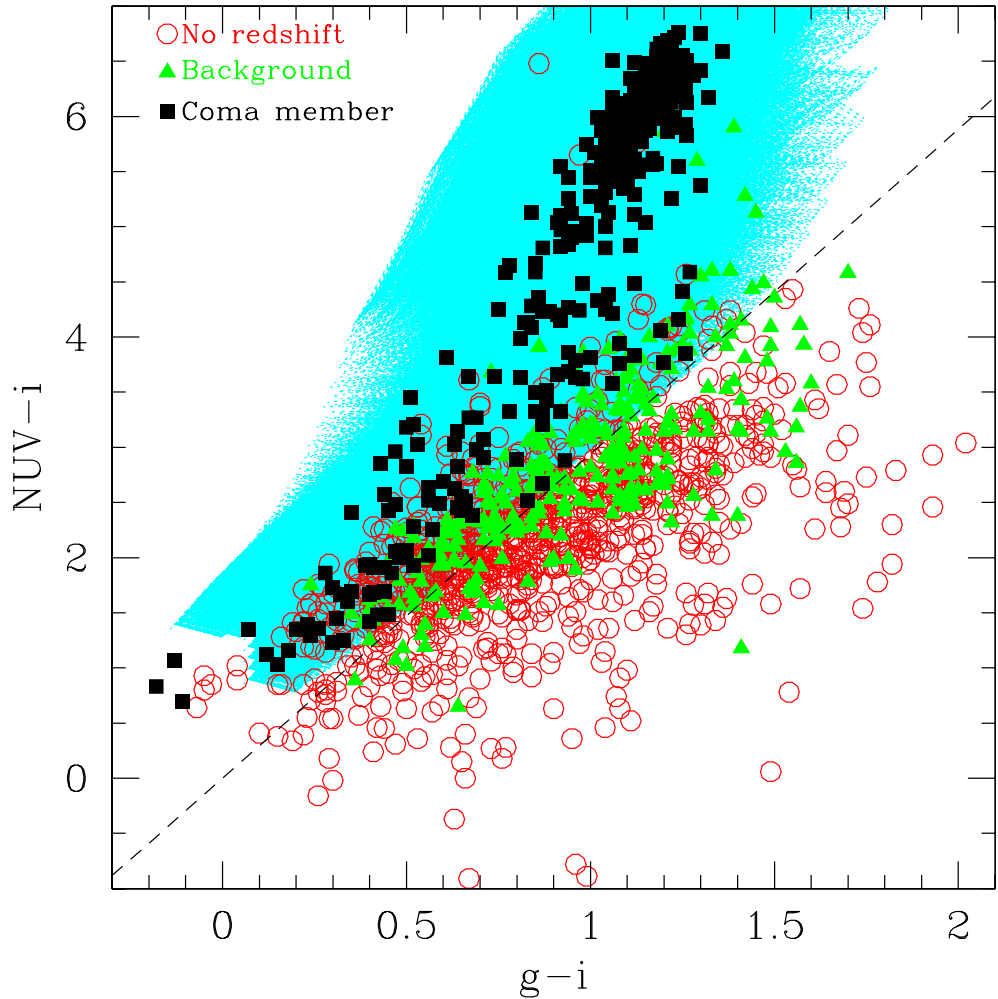}
\caption{\label{colourcut}. The $g-i$ vs. $NUV-i$ observed colour-colour plot used to separate possible members and background galaxies in our sample. Confirmed Coma cluster members, background/foreground objects and galaxies without redshift 
information are shown with squares, triangles and circles respectively. The cyan lines indicate the tracks obtained from 
the SED library of Cortese et al.(2008) assuming an 'a la Sandage' SFH (Gavazzi et al. 2002a), a galaxy age of 13 Gyr, a range of $\tau$ (the time at which the star formation rate reaches the highest value over the whole galaxy history) between 0.1 and 25 Gyr, metallicities between 0.02 $\leq Z \leq$ 2.5 Z$_{\odot}$ and dust attenuation in the range 0$<A(FUV)<$4. 
The dashed line is the threshold adopted to exclude probable background galaxies 
from our sample.}
\end{figure}

The determination of the cluster UV LF requires a reliable estimate of the contribution from 
background/foreground objects to the UV counts. 
This can be accurately obtained for $m_{AB}\leq$18.0, since at this limit our redshift completeness is $\geq$90\% (see Fig.~\ref{completeness}).
 The redshift completeness drops rapidly at fainter magnitudes, thus requiring a statistical estimate of the contamination. 
Three different methods have been here adopted for the computation of the cluster LFs. 
The first one is based on the statistical subtraction of field galaxies, per bin of UV magnitude, that are expected to be randomly projected onto the cluster area, as derived by \cite{xu}. 
In the second one, cluster members are identified on morphological grounds. 
Although sometimes subjective, in particular in the case of faint and/or compact objects, this technique 
can be applied to nearby clusters of galaxies when the redshift coverage is inadequate (e.g. \citealp{sandage85,ferguson88,tully02}). 
To do so, we have visually inspected SDSS-RGB images and we have used the criteria discussed in \cite{vcc} to discriminate between Coma 
and background galaxies. In ambiguous cases, we adopted the colour cut described below to select possible cluster members.
The third one is the completeness-corrected method proposed by \cite{depropris03}.
 This method is based on the assumption that the spectroscopic sample (membership confirmed spectroscopically) is 'representative' of the entire cluster. That is, the fraction of galaxies that are cluster members is the same in the (incomplete) spectroscopic sample as in the (complete) photometric one. 
However, this is not always the case. 
For example, redshift estimates of galaxies in the Coma region are obtained from different sources and the selection criteria are not always well defined. Moreover, spectroscopic targets are often selected according to their colour (e.g. \citealp{mobasher}) or narrow-band emission (e.g. \citealp{jorge02}) in order to maximize the number 
of members in the final sample, thus overestimating the real fraction of members.
It is therefore likely that the membership fraction in our sample represents an upper limit to the real value (in particular at faint apparent magnitudes where the redshift completeness and the fraction of members are low).
The only way to reduce this bias is to try to increase the completeness of our sample by excluding those galaxies which are likely 
to be background objects. We thus used stellar population synthesis models and a ($NUV-i$) vs. ($g-i$) colour 
diagram to reject galaxies with colours not representative of the Coma members (4000 \kms $< V <$ 10000 \kms).
In Fig.~\ref{colourcut} we compare the distributions of Coma confirmed members (squares), background galaxies (triangles) and galaxies 
without redshift estimate (circles) with the spectral energy distribution (SED) library typical of nearby galaxies described in 
\cite{afuv_luca}, whose models span a wide range of ages, metallicities and dust attenuations. 
Every confirmed Coma cluster member lies within the parameter space admitted by the models, allowing us to reduce the number of possible members in our sample. We excluded all galaxies with $(g-i)<$0.34$(NUV-i)$ (dashed line in Fig.~\ref{colourcut}), reducing our sample from 1640 to 853 galaxies in NUV and from 762 to 500 galaxies in FUV.
However, the improvement in the redshift completeness for our sample appears significant only in the last bin, as shown by the dotted and dashed lines in Fig.~\ref{completeness}.

In addition, it is important to remember that the spectroscopic sample is optically selected whereas the photometric sample is UV selected. 
Therefore, objects bright in optical but faint in UV, like elliptical galaxies, have a completeness distribution significantly different from that of spiral galaxies (see Fig.~\ref{completeness}, middle and bottom panel).
Thus we decided to modify the completeness-corrected method 
and to treat separately quiescent (Q) and star forming (SF) galaxies. In particular, we used an observed colour $NUV-r$= 4.5 mag (which corresponds to a specific star formation rate SSFR$\sim$10$^{-11.5}$ yr$^{-1}$) to separate blue/star-forming galaxies from quiescent objects.
For each type and magnitude bin $i$, we then counted the number of cluster members $N_{M}$ (i.e., galaxies with velocity in the range 4000 \kms $< V <$ 10000 \kms), the number of galaxies with a measured recessional velocity $N_{Z}$, and the total number of galaxies $N_{T}$. The completeness-corrected number of cluster members for each type in each bin is: 
\begin{equation}
N_{i} (SF, Q) =  \frac{N_{M} (SF, Q)\times N_{T} (SF, Q)}{N_{Z} (SF, Q)}.
\end{equation}
$N_{T}$ is a Poisson variable, and $N_{M}$ is a binomial variable (the number of successes in $N_{Z}$ trials with probability $N_{M}/N_{Z}$) and the errors associated with $N_{i} (SF, Q) $ are given by:
\begin{equation}
\big(\frac{\delta N_{i} (SF, Q) }{N_{i} (SF, Q) }\big)^{2} = \frac{1}{N_{T} (SF, Q) } + \frac{1}{N_{M} (SF, Q) } - \frac{1}{N_{Z} (SF, Q) }. 
 \end{equation}
 The total number of galaxies in each bin of the LF and its uncertainty are therefore:
 \begin{equation}
N_{i} = N_{i}(SF) + N_{i}(Q), ~~~ \delta N_{i} = \sqrt{\delta N_{i}(SF)^2 + \delta N_{i}(Q)^2}
\end{equation}

\begin{figure*}
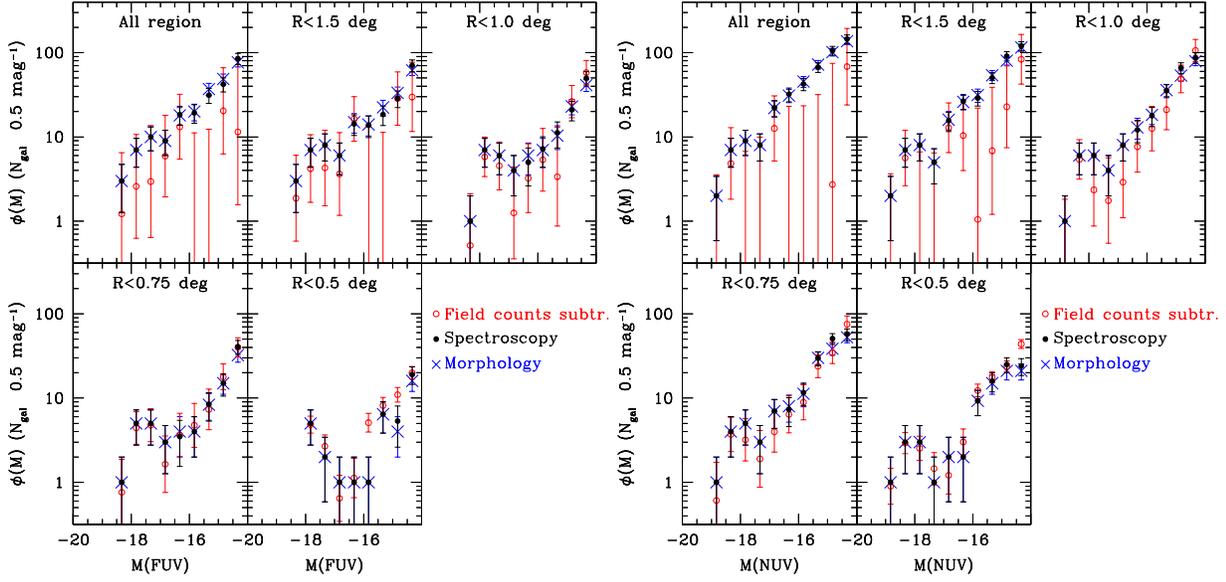

\centering
\includegraphics[width=8.cm]{LF_FUV_testmethods.epsi}
\includegraphics[width=8.cm]{LF_NUV_testmethods.epsi}
\caption{\label{testmethods}The FUV (left) and NUV (right) LFs of the Coma cluster within 
different circular apertures.
Empty and filled circles show the LFs obtained using the background subtraction technique and the spectroscopic completeness method respectively.
The crosses indicate the LFs obtained when cluster members are identified on morphological grounds.}
\end{figure*}

The NUV and FUV LFs obtained using the three methods described above, within different circular apertures centered on the cluster center, are shown in Fig.~\ref{testmethods}. 
All methods show a reasonable agreement for a cluster-centric distance $R<$1 deg (1.67 Mpc) whereas for larger areas the statistical subtraction of the field provides 
number counts significantly lower than the other two techniques. 
This is likely due to the fact that, for scales larger than $\sim$1 deg, the cluster overdensity in UV starts to disappear.
No significant difference is observed between the completeness-corrected and the morphology-based methods.
For these reasons, in the following we will measure the UV LF of Coma using the completeness corrected method only\footnote{All the results presented in the following 
do not change if the background subtraction (for $R<$1 deg) or the morphology are used to estimate the LFs.}.

\begin{figure*}
\centering
\includegraphics[width=14cm]{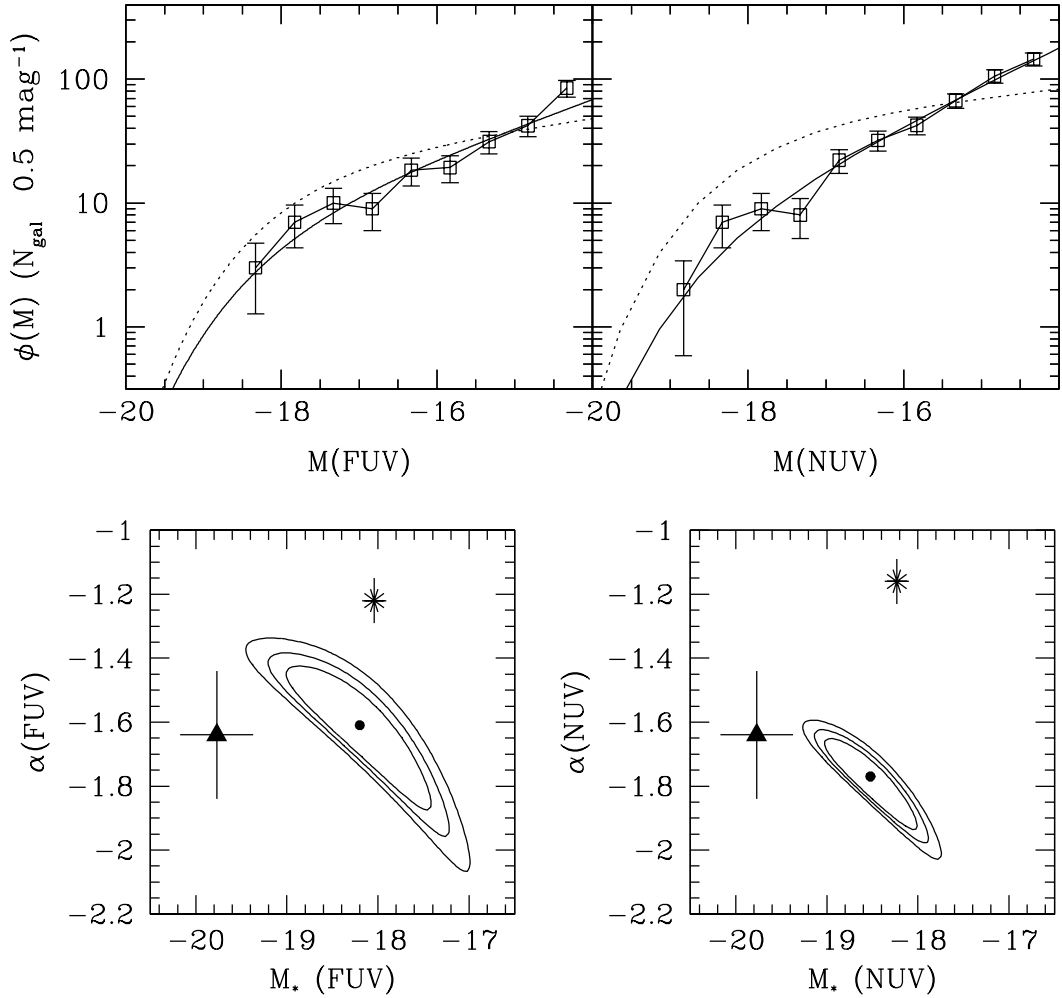}
\caption{\label{allregion}Upper panels: The FUV(left) and NUV (right) LFs for the whole $\sim$9 deg$^{2}$ observed by GALEX in the Coma cluster. The dotted lines show the GALEX LFs for local field galaxies as obtained by Wyder et al.(2005). The field LFs have been normalized in order to match the total number of galaxies detected for M$\leq-$14 mag. The solid lines 
indicate the best-fitting Schecther functions to the data. Lower panels: Contour plots of the 68\%, 95\% and 99\% confidence 
levels for the values of $M_{*}$ and $\alpha$. The results obtained for the field LF and for Abell1367 (Cortese et al. 2005) are indicated with an asterisk and a triangle respectively.}
\end{figure*}

The NUV and FUV LFs obtained for the whole $\sim$9 deg$^{2}$ here investigated are 
shown in Fig.~\ref{allregion} (upper panels). The LFs are not well described by a Schechter function \citep{schechter}, in particular for $M(NUV)<-$17 mag. However, in order to compare our results with previous determinations of field and cluster UV LFs, we fitted our points with a Schechter function by minimizing the $\chi^{2}$.
The best-fit values for $M_{*}$ and $\alpha$ so obtained are presented 
in Table~\ref{bestfit} and compared with previous determinations for local field and cluster galaxies.  
The $\chi^{2}$ contours are shown in the bottom panels of Fig.~\ref{allregion}.
The faint end slopes of the Coma LFs are consistent with those recently obtained by \cite{CORBGA05} in Abell1367 
using GALEX observations (see Table~\ref{bestfit}) and with previous determination of the UV LF in nearby clusters based on balloon-borne experiments \citep{COGA03,redshift}.
On the contrary, Coma shows a bright end $\sim$ 1.5 mag fainter than the one observed in Abell1367. This apparently remarkable difference is due to the presence in Abell1367 of a few galaxies with enhanced star formation, among 
which the ultra-luminous UV galaxy UGC6697 \citep{ugc6697,atlas2006}. Excluding the brightest point 
in the LF of Abell1367 we find a good agreement between the LFs of the two clusters.
When compared to the field (\citealp{wider}, see Table~\ref{bestfit}), the Coma UV LFs have approximately the same bright end ($M_{*}$) but show a considerably 
steeper faint-end slope ($\alpha$), in particular in the NUV (see Fig.~\ref{allregion}).
By dividing our sample into star-forming and quiescent galaxies according to their observed $NUV-r$ colour (see Fig.~\ref{allregiontype}, filled symbols), it clearly appears that the difference between field and cluster LFs is mainly due to the significant contribution of quiescent systems at low UV luminosities. These are not only elliptical and lenticular galaxies (e.g. \citealp{bosell05,donas06}), but also passive spirals whose star formation has been recently quenched by the cluster environment (e.g. \citealp{n4569,dEale}), as shown in Fig.~\ref{mosaic}.
In fact, if we use morphology instead of colour to separate late and early type galaxies (shaded and dotted regions in Fig.~\ref{allregiontype}), 
the number of red galaxies contributing to the faint end of the LF decreases.
The presence of passive spirals is mainly evident in NUV: e.g. for $M(NUV)>-$16 mag  
nearly half of the cluster UV emitting objects are not forming stars, $\sim$70\% early types and $\sim$30\% quiescent 
late type galaxies.
This is probably due to the different stellar populations responsible for the NUV and FUV emission in quiescent systems \citep{bosell05,donas06}.
On the contrary, it is interesting to note that the UV LFs for Coma blue/star-forming galaxies are not significantly different from that of the field.
We can therefore conclude that (1) the faint end slope of the UV LFs are steeper for Coma than for the local field and (2) this difference is almost entirely due to the contribution of red/quiescent galaxies  
at low UV luminosities, consistently with previous studies \citep{COGA03,CORBGA05}.

Finally, we notice that the lack of GALEX observations in the central $\sim$0.26 deg$^{2}$ of Coma, 
mentioned in \S~2, does not affect our conclusions. 
In fact, \cite{redshift} used UV (2000 \AA) FOCA \citep{foca} observations to determine the UV LF of the central $\sim$0.7 deg$^{2}$ of 
Coma for $M_{*}(2000{\rm \AA})<-$15.5 mag. They find a faint-end slope ($\alpha\sim-$1.65$\pm$0.30), consistent with our results. Moreover, \cite{COGA03} combined FOCA observations of the central part of Coma and Abell1367 with FAUST \citep{faust} observations of the Virgo cluster and determined a composite 
local UV LF for galaxy clusters, finding a shape of the LF very similar to the one obtained in this work ($M_{*}=-$18.79$\pm$0.40 mag, $\alpha\sim-$1.50$\pm$0.10).

\section{The radial variation of the UV luminosity functions}
\begin{figure}
\centering
\includegraphics[width=14cm]{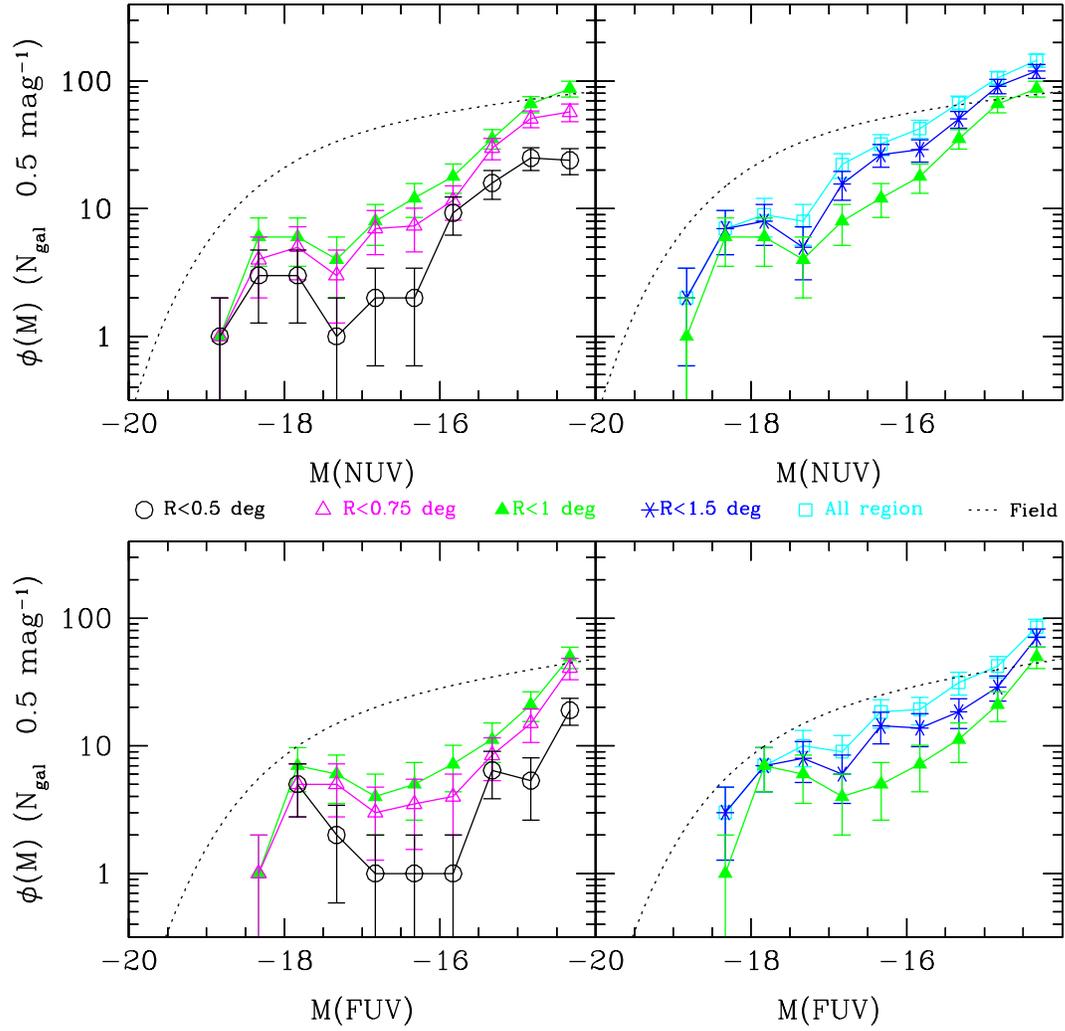}
\caption{\label{LFradius}The NUV (upper panel) and FUV (bottom panel) LFs of the Coma cluster within different circular apertures: 0.5 (circles), 0.75 (empty triangles), 1 (filled triangles), 1.5 deg (stars) and the whole region (squares). The dotted line shows the local field LF, normalized in order to match the total number of cluster galaxies brighter than $M=-$14 mag.}
\end{figure}

The wide area surveyed by GALEX can be used to investigate 
the variation in the shape of the LF as a function of the cluster-centric distance.
Ideally, this should be done by comparing the LF obtained in different circular {\it annuli} centered 
on the cluster center.
In reality, this is not always possible given the poor statistics of current samples. 
For example,  in our case the small number of galaxies at high UV luminosities and the uncertainty in the membership fraction at low luminosities make impossible any estimate of the differential LF.
Any difference observed could in fact only reflect statistical fluctuations in our sample.
For these reasons, we decided to investigate the spatial variation of the LFs focusing our attention on the integral UV LFs, i.e. calculated within different cluster-centric {\it apertures}. 
Inevitably, by using this method, any difference in the shape of the LFs will be less evident 
since the central part of the cluster will contribute to all the LFs. 
The FUV and NUV LFs obtained within circular regions having radius 0.5, 0.75, 1, 1.5 deg ($\sim$0.84, 1.25, 1.67, 2.5 Mpc or 
$\sim$0.3, 0.4, 0.6, 0.9 virial radii) and for the whole sample  are shown in Fig.~\ref{LFradius}.
At all radii the LFs in both NUV and FUV are inconsistent with a Schecther function.
Our data are better described by a Gaussian at bright luminosities and a power law at faint luminosities, complicating the quantification of any difference between LFs at various radii. 
As discussed in the previous section, this is the result of the different contribution of 
star-forming objects and quiescent galaxies to the UV LFs.
Excluding a variation in the normalization, no significant 
change is observed in the shape for different cluster-centric radii. 
The apparent flattening in the faint-end slope at smaller radii marginally visible in NUV is likely due to statistical fluctuations.
We notice that a fit of the faint end slope in the range $-$16$\leq M(UV)\leq-$14 mag would result in a 
significant steepening with increasing radius in NUV (from $\alpha\sim -$1.2 to $\sim-$1.8) and flattening 
with increasing radius in FUV (from  $\alpha\sim-$2.5 to $\sim-$1.7). 
This points  out the danger of blindly fitting faint-end slopes to compare the shape of the LFs obtained with different samples.

\begin{figure}
\centering
\includegraphics[width=14cm]{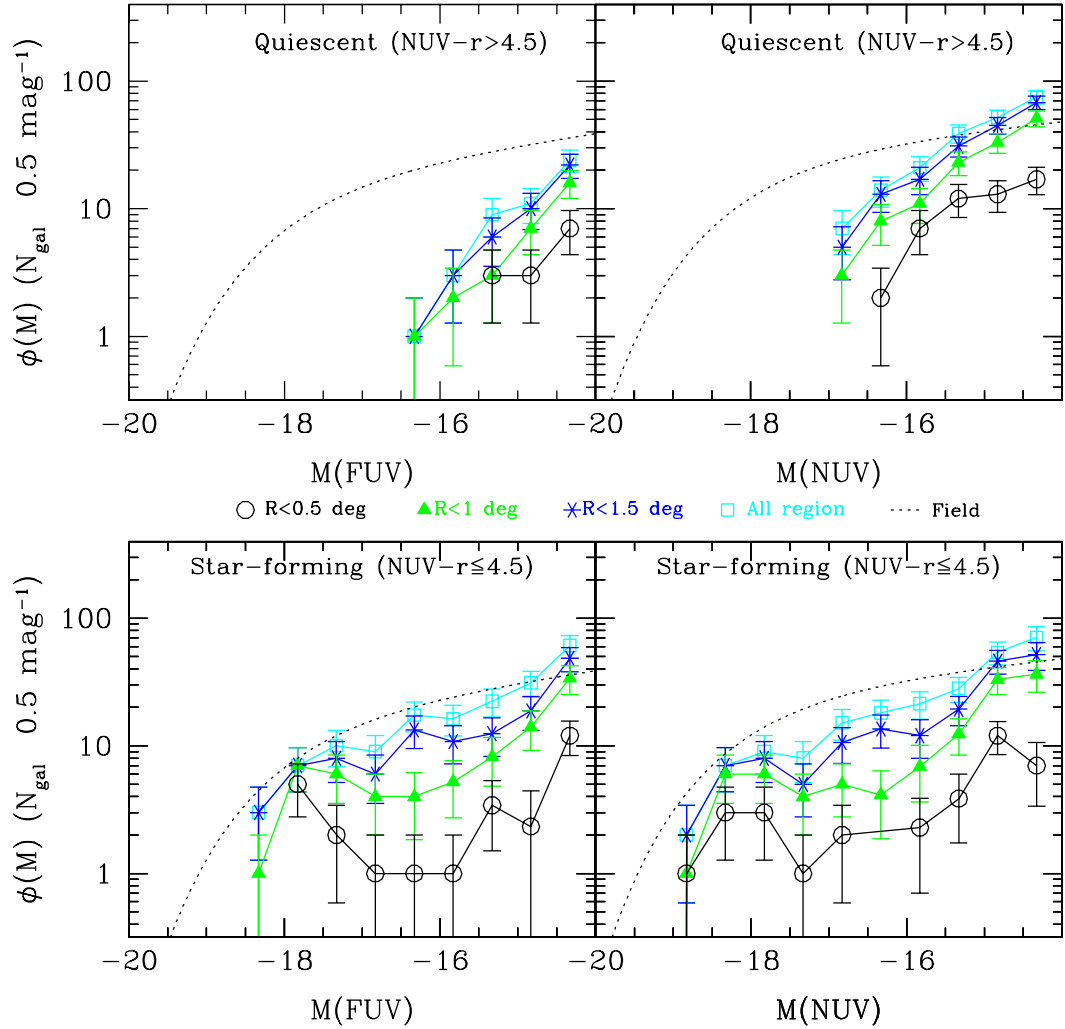}
\caption{\label{LFtypesradius}The NUV (right) and FUV (left) LFs for quiescent systems (upper panel) and star-forming galaxies (bottom panel) within different circular apertures. Symbols are as in Fig.~\ref{LFradius}.}
\end{figure}

In Fig.~\ref{LFtypesradius} we show the integrated NUV and FUV LF as a function of the cluster-centric distance for star forming and quiescent objects separately.
Also in this case we find that, within statistical errors, the shape of the LFs does not vary with cluster-centric distance.
Similar results are obtained if we divide our sample in late and early types accordingly to their morphology (not shown).
Only by integrating the LFs in the range in the range $-$17$<M(NUV)<-$14 mag ($-$16.5$<M(FUV)<-$14 mag), we find that the ratio of the UV emission of star-forming to that of quiescent systems 
monotonically increases from the center ($R<$0.5 deg) to the outskirts,  as expected from the star formation-density relation.
This ratio increases from  $\sim$0.6$\pm$0.3 to $\sim$1.2$\pm$0.2 in NUV and from 
 $\sim$1.8$\pm$1.0  to $\sim$4.1$\pm$0.8 in FUV.
 
Our results are consistent with the recent analysis carried out by \cite{popesso06} on RASS-SDSS galaxy clusters who find 
that the shape of the LFs of blue galaxies and bright red objects does not vary significantly with the distance from 
the cluster center. They only find a significant steepening at large radii in the faint-end slope of the LF of red dwarf galaxies which,
being extremely faint in UV, are not included in our sample.
Also \cite{mobasher}, by studying the $r$ and B band LF of the central region of Coma do not find a significant 
spatial variation in the shape of the LF.
On the contrary, \cite{beije02} find a significant steepening of the LF at larger radii. This effect is more pronounced 
in $U$ band and it has been interpreted as evidence of an infalling population of dwarf galaxies in the outskirts of Coma.
However the variation in the faint-slope mainly occurs in the central $\sim$0.3 deg, where our UV selected sample is 
highly incomplete and cannot be used to test this result. Similarly, the spatial variation in the faint 
end slope of the Coma optical \citep{adami07} and far-infrared \citep{bai_coma} LFs cannot be tested because 
obtained for regions (e.g. the core of Coma) or magnitude ranges ($R>$21 mag) not included in our UV selected sample.

We can therefore conclude that, at least in the range 0.5$\leq R \leq$ 2 deg, the shape of the integrated NUV and 
FUV LFs of the Coma cluster remains constant. We remind the reader that our results are only valid for integrated LFs and that the small number statistic makes impossible any quantification of the shape of the differential UV LFs.

\section{The extinction corrected LF and SFR of 
the Coma cluster}

\begin{figure*}
\centering
\includegraphics[width=14cm]{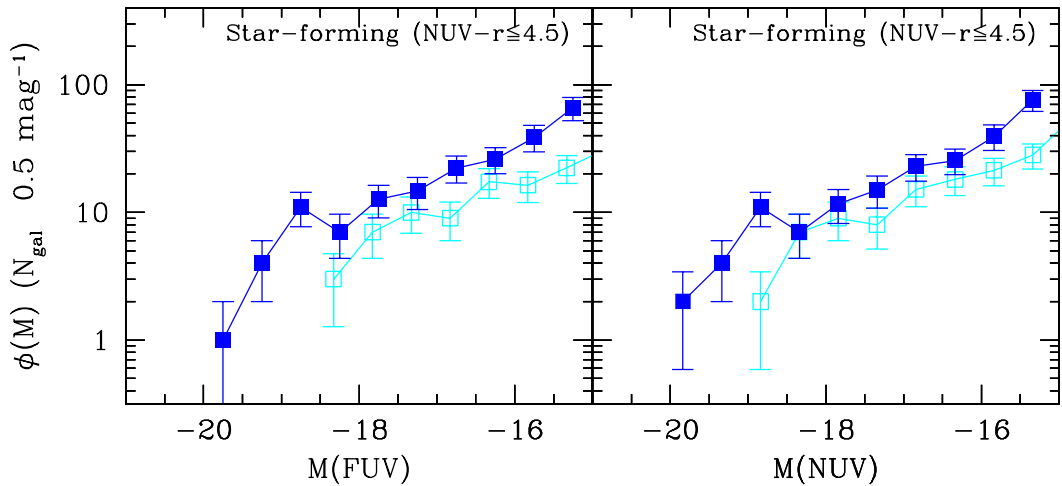}
\caption{\label{LFdust} Effect of dust attenuation corrections on FUV (left) and NUV (right) LFs of star-forming galaxies.  Empty and filled symbols show the LFs before and after internal extinction 
correction respectively.}
\end{figure*}

Previous determinations of UV LFs in all environments and at all redshifts 
have never taken internal extinction corrections in account.
This is mainly due to the fact that methods for estimating dust attenuation usually require far-infrared photometry, quite rare for 
large samples, and are calibrated only on starbursts and active star-forming galaxies.
Only recently, \cite{COdust05,afuv_luca} have provided UV dust attenuation correction recipes, valid 
for galaxies with different star formation histories. These recipes only require two broad-band colours and can 
be used to estimate an UV LF corrected for dust attenuation and to determine the total SFR in Coma.
Since UV dust attenuation is important only for star-forming galaxies, in the following we restrict our analysis to the 
subsample including only blue galaxies (i.e. $NUV-r\leq$4.5).
For each galaxy, we determined $A(FUV)$ and $A(NUV)$ using the observed $FUV-NUV$ and 
$NUV-i$ colours as described in \cite{afuv_luca}\footnote{The attenuations in the two GALEX bands have been obtained 
independently, without any assumption on the shape of the extinction curve.}.
We then corrected the NUV and FUV magnitudes and determined the NUV and FUV LF using the 
completeness corrected method described in \S 3. 
The LFs obtained for the whole sample are shown in Fig.~\ref{LFdust}: after dust attenuation correction, the LFs shift to brighter luminosity by $\sim$1.5 mag in FUV and $\sim$1 mag in NUV, but their shapes do not change significantly.
We also determined the unreddened LFs within different apertures and found no significant radial variation 
in their shapes (not shown).

By integrating the observed and unreddened LFs within the magnitude range investigated here ($M\leq-14$ mag), we 
determined the fraction of UV luminosity absorbed by dust. The total UV dust attenuations of the Coma cluster are $A(NUV)$= 0.84$\pm$0.28 mag and $A(FUV)$=1.25$\pm$0.25 mag, implying that two third  (half) of the light emitted in 
FUV (NUV) by Coma galaxies is absorbed by dust. No variation in the UV dust attenuation with the distance from the cluster 
center is observed. These values are consistent with the total FUV dust attenuation observed in the local field 
$A(FUV)$=1.29$^{+0.32}_{-0.30}$ mag \citep{afuv_field}, suggesting that the dust properties of  blue/star-forming galaxies in Coma are not significantly different from those of field star-forming systems. 

The unreddened UV LFs of star-forming galaxies can be also used to estimate the total SFR of the Coma cluster.
To do so we integrated the LFs in the observed range and then converted the total luminosity 
into total SFR following \cite{jorge06}:
\begin{eqnarray}
\label{sfr}
SFR(FUV)~[{\rm M_{\odot}~yr^{-1}}]= \frac{L(FUV) [{\rm erg~sec^{-1}}]}{3.83\times10^{33}}    \times 10^{-9.51}    \nonumber\\
SFR(NUV)~[{\rm M_{\odot}~yr^{-1}}]= \frac{L(NUV) [{\rm erg~sec^{-1}}]}{3.83\times10^{33}}    \times 10^{-9.33}    
\end{eqnarray}
where $L(NUV)$ and $L(FUV)$ are the total FUV and NUV luminosities (corrected for dust attenuation) of the Coma cluster\footnote{We note that, in reality, the calibration constant for the UV luminosity vs. SFR relation 
varies by up to a factor $\sim$1.5 with the stellar metallicity: i.e. the UV luminosity increases at lower metallicities  \citep{bicker05}. Unfortunately we do not 
have accurate metallicity information to take this effect into account.}.
The total SFR of the Coma cluster results\footnote{We consider the average between the total SFR obtained in NUV and FUV}  $SFR$=90$\pm$16 M$_{\odot}$ yr$^{-1}$, consistent with the $SFR(FIR)$= 97 M$_{\odot}$ yr$^{-1}$ recently obtained by \cite{bai_coma} using far-infrared observations.
A similar value is obtained if the total SFR is estimated combining the FIR and UV luminosities \citep{bell03b,hirashita03,jorge04b}:
\begin{equation}
SFR = SFR^{0}_{NUV} + (1-\eta)SFR(FIR) 
\end{equation}
where $\eta$ accounts for the IR cirrus emission (here we assume $\eta$=0.3, \citealp{bell03b}) and $SFR^{0}_{NUV}$ is obtained from equation (\ref{sfr}) but using the observed NUV luminosity.

\begin{figure}
\centering
\includegraphics[width=14cm]{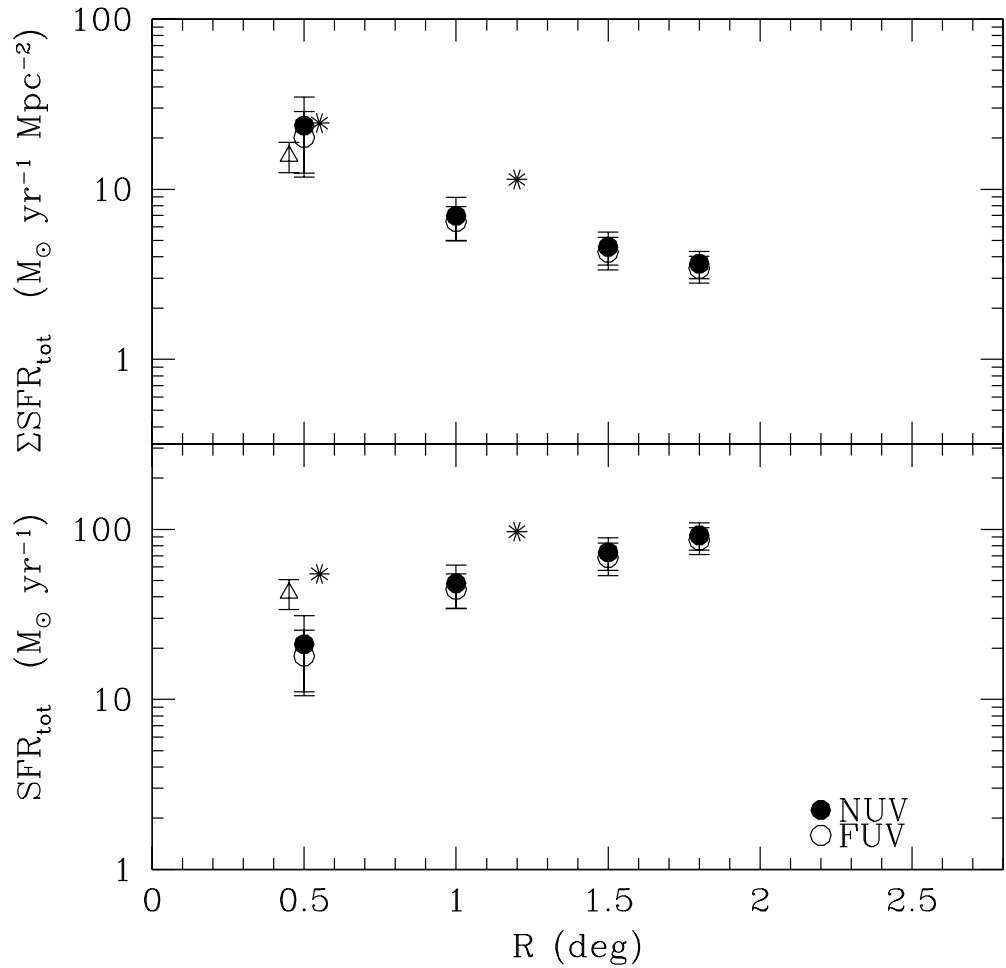}
\caption{\label{sfrradius} The SFR surface density (upper panel) and integrated SFR of the Coma cluster (bottom panel) as a function of the cluster-centric distance. Filled and empty circles show the SFR obtained from NUV and FUV luminosities respectively. Asterisks  and triangles indicate 
the values recently obtained by Bai et al.(2006) using MIPS observations and 
by Iglesias-Paramo et al.(2002) using H$\alpha$ narrow-band imaging respectively.}
\end{figure}

This result suggests that totally obscured star formation (i.e. not at all visible in UV), if present, represents a very small 
fraction of the total star formation in Coma, in agreement with IRAS (e.g. \citealp{bicay87,review}) and Spitzer (e.g. \citealp{muzzin08,saintonge08}) FIR observations of local clusters of galaxies.\footnote{In the near future, when Spitzer and Herschel data will become publicly available, it will also be possible to test this 
result on a galaxy by galaxy basis.} 
In Fig.~\ref{sfrradius} we show the variation of the integrated SFR and SFR surface density as a function of cluster-centric distance and we compare our results with previous estimates based on 
far-infrared observations  \citep{bai_coma} and H$\alpha$ narrow-band imaging \citep{jorge02}. As expected, the integrated SFR increases by a factor $\sim$4, and the integrated SFR surface density decreases by a factor $\sim$5, from the center to the outskirts of Coma.
Although the total SFR obtained integrating over the whole area observed by GALEX is roughly consistent with the value obtained from far-infrared observations, the SFR obtained in UV 
for $R<$1 deg is $\sim$2-3 times lower than that obtained in similar areas from far-infrared and H$\alpha$ data.
This difference is likely due to the absence of GALEX observations in the central $\sim$0.26 deg$^{2}$, as supported by the similar SFR surface density obtained at small radii in UV, far-infrared and H$\alpha$ (see Fig.~\ref{sfrradius}).

\begin{figure}
\centering
\includegraphics[width=14cm]{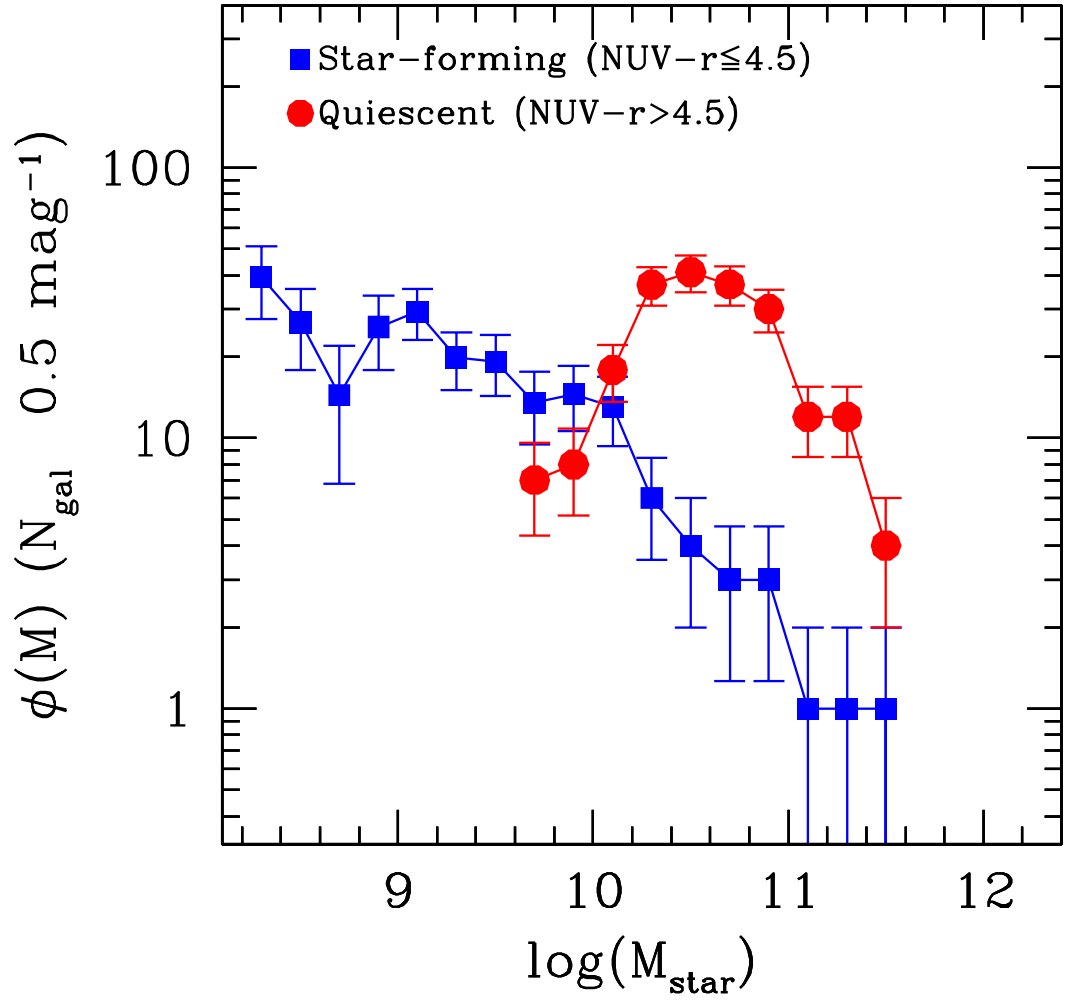}
\caption{\label{massfunc} The stellar mass distributions for quiescent (circles) and star-forming galaxies (squares) in our UV-selected sample.}
\end{figure}

The contribution of Coma-like clusters of galaxies to the SFR density in the local universe can be computed if the cluster 
volume density is known. This value varies significantly with the properties of the adopted cluster sample (e.g. \citealp{scaramella91,mazure96,bramel00,depropris02}), but we can roughly assume it to lie in the range 3-9.5 $\times$ 10$^{-6}$ Mpc$^{-3}$.
Thus, the total SFR density in local clusters is $\rho_{SFR}(cl)\sim$2.5-8.5 $\times$ 10$^{-4}$ M$_{\odot}$ yr$^{-1}$ Mpc$^{-3}$, i.e. between 1.5\% and 5\% 
of the total SFR density in the local universe $\rho_{SFR}\sim$ 10$^{-1.80\pm0.16}$ M$_{\odot}$ yr$^{-1}$ Mpc$^{-3}$ \citep{hanish06}.
This value mainly depends on the density of local clusters and does not provide any information on the different SFH of cluster and field galaxies.
A more interesting exercise is the comparison of the SSFR of Coma and the field.
In order to quantify the total mass for our sample, we converted the unreddened $i$ band magnitude to stellar mass using the 
$g-i$ colour following \cite{bell03} and estimated the stellar mass distribution by using the completeness-corrected method described in \S ~3.
The stellar mass distributions for the star-forming and quiescent galaxies in the UV-selected sample are shown in Fig.~\ref{massfunc}.
The total stellar mass of our sample results $M_{star}(total)$=10$^{13.13\pm0.10}$ M$_{\odot}$,  great part of which 
is assembled in quiescent systems: i.e.  $M_{star}(Q)$=10$^{13.07\pm0.09}$ M$_{\odot}$ and $M_{star}(SF)$=10$^{12.26\pm0.20}$ M$_{\odot}$.
In Fig.~\ref{massfunc}, it clearly emerges that the vast majority of the star forming galaxies are low mass ($\leq10^{10}$ M$_{\odot}$) 
objects. Moreover, by estimating the UV LF for low and high mass objects it emerges that $\sim$66$\pm$10\% of the
total Coma SFR is taking place in galaxies less massive than $\sim10^{10}$ M$_{\odot}$, confirming previous 
observational evidences that \emph{downsizing} is also present in galaxy clusters (e.g. \citealp{boselli,review,delucia04_cl,phenomen,gav02,kodama04,poggianticoma,poggianti06}).
The integrated SSFR for the Coma cluster is $\sim$ 10$^{-11.18\pm0.13}$ yr$^{-1}$, significantly lower than the the typical SSFR observed in the local field e.g. $SSFR_{field}\sim$ 10$^{-10.17}$ yr$^{-1}$ \citep{salim07} or 
$SSFR_{field}\sim$ 10$^{-10.55\pm0.20}$ yr$^{-1}$ if we assume $\rho_{SFR}\sim$ 10$^{-1.80\pm0.16}$ M$_{\odot}$ yr$^{-1}$ Mpc$^{-3}$ \citep{hanish06} and a local stellar mass density $\sim$10$^{8.75\pm0.12}$ M$_{\odot}$ Mpc$^{-3}$ \citep{perez08}.
This difference is likely a combination of the morphology-density and star formation-density relations: massive elliptical galaxies are more overdense in clusters \citep{dressler80,whitmore} but also, for the same morphological type, cluster galaxies have a lower star formation activity than field objects \citep{ha06}. 
Finally, it is interesting to note that the SSFR of blue/star-forming galaxies is $\sim$ 10$^{-10.31\pm0.22}$ yr$^{-1}$, i.e. consistent with that of the field, suggesting that Coma star-forming galaxies have not been significantly affected by the cluster environment.

\subsection{Bias and selection effects}
The results presented above might be affected by two different selection effects. Firstly, we 
are underestimating the total SFR of Coma since we lack observations for the cluster core. 
Secondly, the selection in UV adopted here is  
strongly biased against red massive systems (e.g. see Fig.~\ref{ssfr}), suggesting that we could 
also underestimate the total stellar mass of Coma.
We can quantify the first bias by comparing our results with previous determinations of the total SFR (\citealp{bai_coma,jorge02}; see Fig.~\ref{sfrradius}) that include the cluster center. 
In the central $\sim$0.5 deg, we are missing $\sim$30 M$_{\odot}$ yr$^{-1}$, implying that the total SFR of Coma presented above is underestimated by a factor $\sim$1.3.
The total SFR density in Coma-like clusters would therefore increase to $\sim$2\%-7\% of the total SFR density in the local universe.
To address the second bias, we can estimate the total stellar mass of the Coma cluster from its total mass (baryonic+dark matter) M$_{tot}\sim$1.4 $\times$ 10$^{15}$ M$_{\odot}$  \citep{lokas03}, assuming a mass-to-light ratio M$_{tot}$/L$_{K}$= 75 M$_{\odot}$/L$_{\odot}$  \citep{rines04} and a stellar mass-to-light ratio M$_{star}$/L$_{K}$ ratio of $\sim$1 M$_{\odot}$/L$_{\odot}$ \citep{bell03}.  We obtain a total stellar mass of 
 $M_{star}\sim$10$^{13.3}$ M$_{\odot}$, i.e. $\sim$1.5 times higher than the value obtained for our sample.
The fact that the underestimates in the total SFR and stellar mass are comparable implies that 
the SSFR of the Coma cluster decreases only by a factor $\sim$1.1, well within the observational errors.
We can therefore conclude that the lack of GALEX observations for the core of Coma does not significantly bias our results.

\section{Discussion}
\subsection{Why are the field and cluster LFs of star-forming galaxies so similar?}
\begin{figure*}
\centering
\includegraphics[width=14cm]{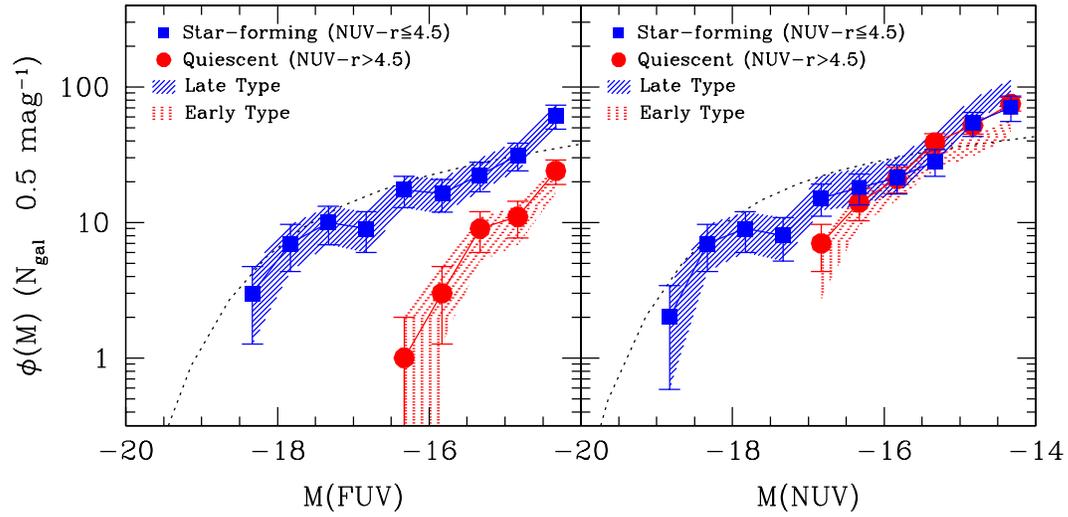}
\caption{\label{allregiontype}The contribution of different galaxy types  to the FUV (left) and NUV (right) LFs.
The LFs of star-forming and quiescent galaxies, classified according to their observed $NUV-r$ colour, are indicated with squares and circles respectively. 
The LFs of late and early types, classified from visual inspection of SDSS images, are the shaded and dotted regions respectively. 
The dotted lines show the GALEX LFs for local field galaxies. The field LFs have been normalized in order to match the total number of cluster 
star-forming galaxies brighter than $M=-$14 mag.}
\end{figure*}

\begin{figure*}
\centering
\includegraphics[width=14cm]{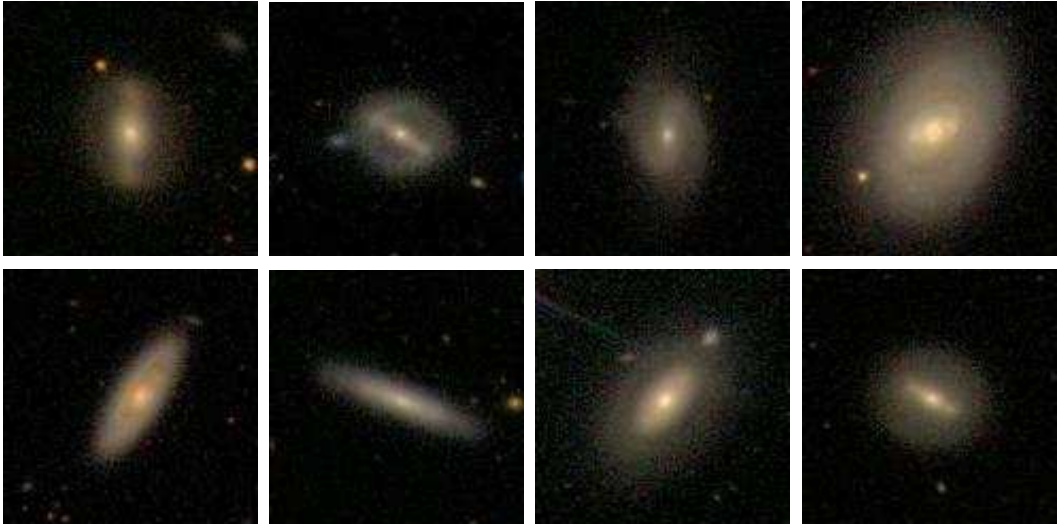}
\caption{\label{mosaic} SDSS colour images of quiescent disk galaxies in 
our UV-selected sample. Each image is 1.2 arcmin per side, corresponding to $\sim$33 kpc at the distance of Coma}.
\end{figure*}

Our analysis of GALEX observations of the Coma cluster provides 
definitive evidence that the shape of the UV LF is not universal 
but depends on the local environment (e.g. see Figs.~\ref{allregion},~\ref{allregiontype}). Coma has a steeper 
faint-end due to the high number density of massive, quiescent galaxies and its 
total SSFR is significantly lower than that of the local field. 
This is 
a direct consequence of the morphology-density and star formation density relations: quiescent galaxies 
are in fact ellipticals, lenticulars and passive spirals whose star formation has been recently quenched 
by the cluster environment (see Fig.~\ref{mosaic}).
Only if we restrict our attention to blue/star-forming cluster galaxies, the UV LFs and SSFR 
are not significantly different from those observed in low density environments,
in agreement with recent studies of nearby (e.g. \citealp{balogh04,CORBGA05}) and high redshift (e.g. \citealp{finn08,cooper08}) clusters.  
These results could be interpreted as a strong evidence that the environment does not significantly affect the cosmic SFH of galaxies \citep{finn08,cooper08}. However, this interpretation is inconsistent with the plethora of evidences showing the 
strong effects of the environment on the evolution of cluster galaxies 
(e.g. see \citealp{review} and references therein). 
Here we show how the apparent independence of the properties of star-forming galaxies from the environment is, in reality, only a selection effect.

\begin{figure}
\centering
\includegraphics[width=14cm]{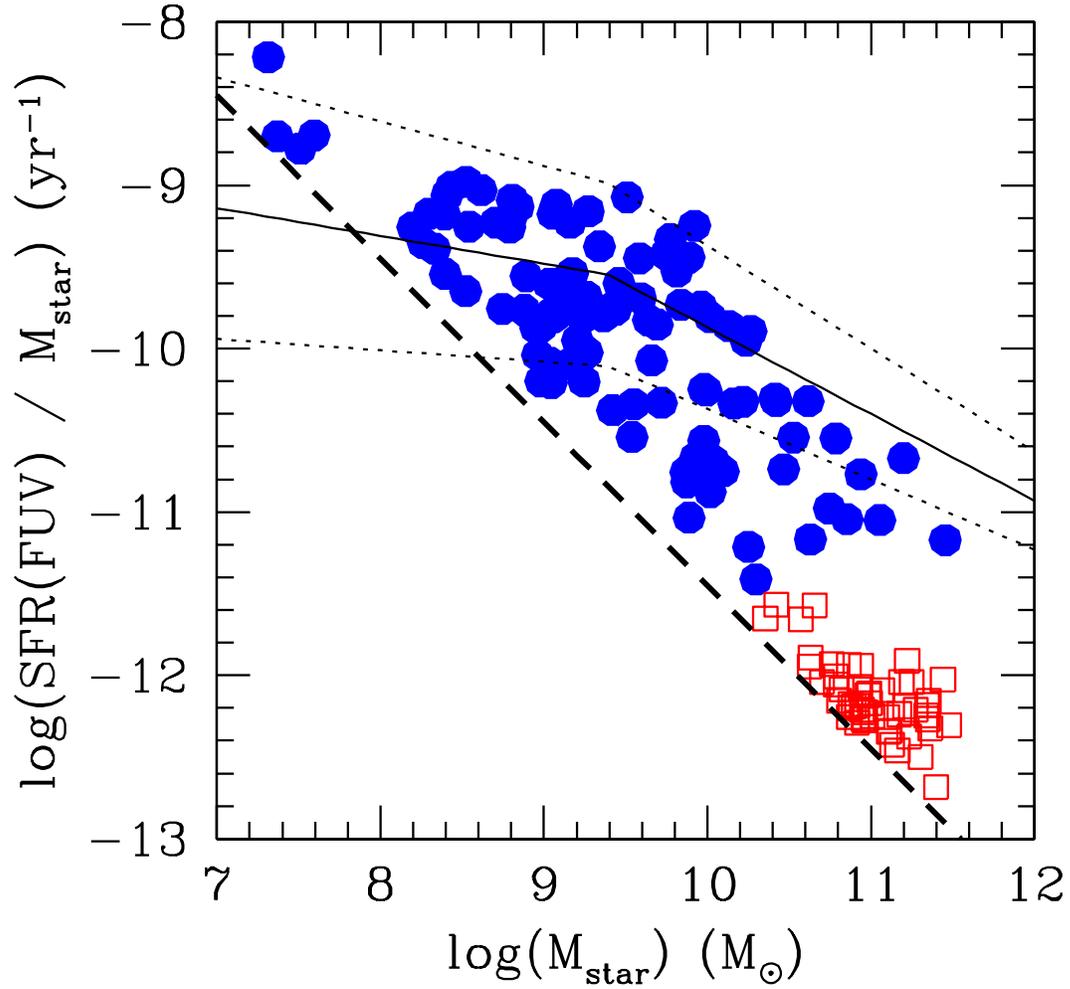}
\caption{\label{ssfr}The SSFR as a function of the stellar mass for Coma confirmed 
members. Circles and squares indicate blue/star-forming and red/quiescent galaxies respectively. The dashed line shows our detection limit in UV. 
The best-fit relation for field galaxies (solid line) and its scatter (dashed line) as obtained by Salim et al. (2007) is superposed.}
\end{figure}

\begin{figure}
\centering
\includegraphics[width=14cm]{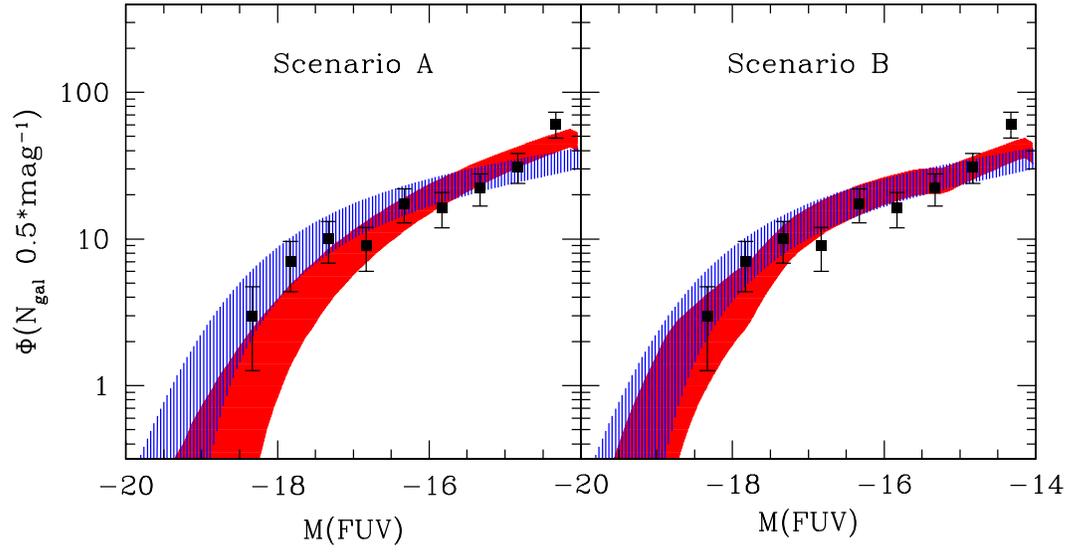}
\caption{\label{simul} Toy model investigating the effects of the environment on the shape of the FUV LF for star-forming galaxies. 
Black squares are the Coma FUV LF for blue/star-forming galaxies in our sample and the blue dashed regions indicate the field FUV LF $\pm$1$\sigma$.
 The red filled areas show the cluster FUV LF $\pm$1$\sigma$ obtained from 
 our simulations assuming a maximum amount of quenching independent of galaxy luminosity (left panel) or decreasing linearly with galaxy luminosity (right panel).}
\end{figure}

Our definition of blue/star-forming galaxies is based on colour or SSFR (i.e. $NUV-r\leq$4.5, SSFR$\geq$10$^{-11.5}$ yr$^{-1}$), as often happens for large samples, 
implying that those galaxies which have been strongly affected by the environment are automatically excluded. This selection effect is evident from Fig.~\ref{ssfr}, where the relation between SSFR and stellar mass for confirmed cluster members in our sample is shown.
The solid line shows the typical relation (and its 1$\sigma$ uncertainty, dotted lines) for field galaxies as obtained by \cite{salim07}, while the dashed line 
represents our sensitivity limit (SFR$\sim$0.04 M$_{\odot}$ yr$^{-1}$). For M$_{star}<$ 10$^{9.5}$ M$_{\odot}$ we detect only healthy star-forming objects and, if a galaxy had its star formation quenched by the environment, it will not be detected by our observations. 
By comparison, for M$_{star}>$ 10$^{9.5}$ M$_{\odot}$, star-forming galaxies have to lie within $\sim$3 $\sigma$ from the field relation, otherwise their $NUV-r$ colour would be too red and they would be classified as quiescent objects.
A clear example is represented by NGC4569 in the Virgo cluster, the prototype 
of anemic spiral. Its star formation activity has been very recently ($<$500 Myr ago) quenched by ram pressure stripping \citep{vollmer04,n4569} and its colour ($NUV-r\sim$4.1 mag, SSFR$\sim$10$^{-11.2}$ Gyr$^{-1}$) is already almost as red as that of a quiescent galaxy.  Passive spiral galaxies are in fact already included in our sample of quiescent 
systems, as shown in Fig.~\ref{allregiontype} and in Fig.~\ref{mosaic}.\\  
In order to demonstrate the validity of this interpretation, we investigated the effects of the 
environment on the shape of the UV LF for star-forming galaxies using a very simple toy model. 
We started with an infalling population of $\sim$1200 galaxies with $M(FUV)<-$14 mag, 
reproducing the field FUV LF.
For each infalling galaxy, the SFR (i.e. $M(FUV)$) is instantaneously 
quenched by the cluster environment. 
The amount of quenching ($SFR_{before}/SFR_{after}$) is randomly chosen, 
assuming a uniform distribution.
Two different scenarii have been considered, depending on the maximum amount 
of quenching allowed. 
In Scenario A, the maximum amount of quenching is $SFR_{before}/SFR_{after}$=100, independently of galaxy luminosity. 
In Scenario B,  the maximum amount of quenching decreases linearly with galaxy luminosity: i.e.  from 
$SFR_{before}/SFR_{after}$=100 for $M(FUV)=-$14 mag  to $SFR_{before}/SFR_{after}$=10 for $M(FUV)=-$18 mag.
By construction, both models are able to reproduce the parameter space occupied by our sample in Fig.~\ref{ssfr}.
For each scenario, we then reconstruct the \emph{cluster} FUV LF including only those galaxies 
still matching our definition of star-forming object, i.e.  $SSFR>$10$^{-11.5}$ yr$^{-1}$ and $SFR>$0.04 M$_{\odot}$ yr$^{-1}$. One thousand simulated LFs have been so obtained and the average result, normalized to the total number of blue galaxies in 
our sample, is shown in Fig.~ \ref{simul}. 
We note that both scenarios are likely to overestimate the real variation of the LF.
In fact, we assume an instantaneous quenching whereas this time-scale is probably longer (see next section). Moreover, a quenching uniformly distributed in the range 1-100 is likely an overestimate of the average decrease in the SFR of cluster galaxies \citep{ha06,gomez03,lewis02}. 
Even so, the variation in the shape of the LF after (filled area in Fig.~\ref{simul}) the infalling into the cluster does not appear very significant\footnote{We remind that we are only interested in the shape of the LF. After the quenching of the star formation the absolute normalization of the LF is significantly reduced.}, supporting our interpretation.
Therefore, even if the cluster is very efficient in quenching 
the SFR, a selection based on colour/SSFR 
automatically excludes those galaxies which have already been strongly affected by the environment.
This selection effect, even more important at higher redshift where the dynamical range in stellar mass 
is usually smaller than ours, implies that cluster and field star-forming galaxies are characterized by a similar LF by construction, since we are 
only selecting healthy objects.
The steepening of the NUV LF for late type galaxies, once quiescent spirals are included (shaded region in Fig.\ref{allregiontype}), provides 
additional support to our interpretation.  
Thus, the important result does not lie in the shape of the LFs but in the fact that we detect a significant fraction of healthy star forming galaxies in the cluster environment.
These galaxies likely represent a recent infalling population of field systems, not yet affected by the harsh environment of Coma. 
This scenario also explains the similar decrease in the SFR of field and cluster galaxies between $z\sim$1 and $z\sim0$ \citep{cooper08,finn08}: the vast majority of cluster star-forming galaxies are still infalling for the first time 
into the cluster center and therefore their properties are still representative of the field.

\subsection{The accretion rate of the Coma cluster}
As argued in the previous section, the healthy star-forming galaxies observed in Coma appear to have just started their first dive into the cluster center.
Our observations could therefore allow us to quantify the mass accretion rate of star-forming galaxies from 
the field assuming that we know the exact time-scale necessary to quench the star formation.
This is still an open issue but, assuming reasonable estimates for the typical Coma crossing time  
($\sim$1.6 $\times$ 10$^{9}$ yr, \citealp{review}), the time-scale associated with the 
various physical processes invoked to suppress star formation ($\sim$0.1-1 Gyr; 
\citealp{ABAM99,review,n4569,dEale,a2667,FUJI04,QUIM00,poggia99,poggianticoma,roediger,shioya02,vollmer01}) and the time-scale 
for the evolution of the UV luminosity ($\sim$10$^{8}$ yr), we conclude that 1-3 Gyr (at most) should be sufficient to transform an infalling galaxy into a quiescent system.  
By combining this time-scale with the total stellar mass of the infalling galaxies we can infer a stellar mass 
accretion rate of $\sim$(0.6-1.8)$\times$10$^{12}$ M$_{\odot}$ Gyr$^{-1}$, roughly consistent with the value obtained by \cite{adami05}, 
when converted to total mass using a total-to-stellar mass ratio M$_{tot}$/M$_{star}$=75 \citep{rines04}.
This implies that, if the mass accretion rate has not significantly changed during the age of the universe, the whole quiescent 
population could have formed by infalling galaxies in a Hubble time.   
In particular, from $z\sim$1 until now the stellar mass in the red sequence has increased by a factor 1.5-5 (depending 
on the real infall rate\footnote{These values are obtained assuming that all the stellar 
mass in the central part of Coma (not included in this work) lies in red sequence. These therefore represent lower limits to the real value.}), a value consistent with the cosmic growth of stellar mass (e.g. \citealp{bell07,brown07,cimatti06,faber07,zucca06}). 
We note that this simple picture is consistent with the large accretion rate in clusters at $z<$1 predicted by numerical 
simulations (e.g. \citealp{delucia04b}) and with the typical infalling rate obtained from observations (e. g. \citealp{ellingson01,andreon06}) and simulations (e.g. \citealp{VdB02,berrier08}). In addition, a steadily growth of Coma over cosmic time would not significantly modify 
the bright end ($M^{*}$) of the cluster stellar mass function, in agreement with observations of high redshift clusters (e.g. \citealp{andreon06b,depropris07}).

The fact that red sequence galaxies are preferentially found in clusters of galaxies 
suggests that the cluster environment could play a non negligible role in the building up of the 
red sequence since $z\sim$1.
If so, we can expect a rapid decrease in the difference between the specific star formation of Coma 
and the field from $z\sim$1 to $z\sim$0.
An accurate quantification of the mass accretion rate of Coma, and of its evolution with redshift, is therefore 
mandatory to determine whether or not the shape of the cosmic SFH depends on the environment.
 
In spite of their large uncertainty, these simple calculations 
show that the infall of healthy spirals from low density environments into the center of cluster 
of galaxies can not only explain part of 
the morphology-density \citep{smith05,desai07} and star formation density relations observed in today clusters of galaxies, 
but could also easily account for the strong evolution in the stellar mass budget of the red sequence from $z\sim$1 to $z\sim$0. 

\section{Summary \& Conclusions}
In this paper, we have presented a study of the UV properties of the Coma cluster based on  
GALEX NUV and FUV observations covering $\sim$9 deg$^{2}$ centered on the cluster core.
Although the central $\sim$0.26 deg$^{2}$ could not be observed, our analysis represents the widest and deepest UV investigation of  a nearby cluster of galaxies made to date.
Our main results are as follows:\\

a) The Coma NUV and FUV LFs show a faint end slope significantly steeper than the 
one observed in the field. This difference is more evident in NUV and it is due to the higher number density of massive 
quiescent/red galaxies (i.e. ellipticals, lenticulars and passive spirals) in Coma compared to the field. 
The contribution of quiescent galaxies to the total UV emission at low luminosities ($M(UV)>-$17 mag) is larger in the cluster center, however no significant variation in the shape of the UV LFs with cluster-centric distance is observed.\\

b) We estimated for the first time the UV LFs of star-forming galaxies corrected for internal dust attenuation. We showed that Coma-like clusters contribute only $<$7\% of the total SFR density of the local universe.
More interestingly the SSFR of Coma is $\sim$ 10$^{-11.18\pm0.13}$ yr$^{-1}$, significantly lower than the integrated SSFR of the local universe.
Approximately 2/3 of the whole SF in Coma is occurring in objects with M$_{star}<$ 10$^{10}$ M$_{star}$, confirming that \emph{downsizing} is also present in high density environments. \\

c) The shape of the UV LF and the SSFR of blue/star-forming galaxies are consistent with those of the field, in agreement with previous works. We have shown that this similarity does not imply that the effects of the environment on the evolution of the cosmic SFH are negligible. On the contrary, these results are consistent with a scenario in which cluster star-forming galaxies are still infalling for the first time into the cluster center. The stellar mass accretion rate of Coma results $\sim$(0.6-1.8)$\times$10$^{12}$ M$_{\odot}$ Gyr$^{-1}$.  At this rate, the whole cluster could have easily formed from infalling galaxies accreted from the field in a Hubble time.
More interestingly, a significant fraction of the population of lenticular and passive spirals observed today in Coma could arise from infalling galaxies accreted between $z\sim$1 and $z\sim$0, perhaps suggesting that the environment plays a significant role in the mass growth of the red sequence in the universe.

\section*{Acknowledgments}
We wish to thank the anonymous referee, whose comments and suggestions were useful for improving this manuscript. 
LC is supported by the UK Science and Technology Facilities Council. 
LC wishes to thank Barbara Catinella for useful discussions and for a critical reading of this manuscript and Jonathan Davies, Tom Hughes, Rory Smith and Rhys Taylor for helpful comments.
GG acknowledges the support by ASI-INAF grant I/023/05/10.
This research is mainly based on the GALEX GI program GALEXGI1-039, as well as public 
archival data, available at the MAST archive.
GALEX (Galaxy Evolution Explorer) is a NASA Small Explorer, launched in April 2003.
We gratefully acknowledge NASA's support for construction, operation,
and science analysis for the GALEX mission,
developed in cooperation with the Centre National d'Etudes Spatiales
of France and the Korean Ministry of Science and Technology.

Funding for the SDSS and SDSS-II has been provided by the Alfred P. Sloan Foundation, the Participating Institutions, the National Science Foundation, the U.S. Department of Energy, the National Aeronautics and Space Administration, the Japanese Monbukagakusho, the Max Planck Society, and the Higher Education Funding Council for England. The SDSS Web Site is http://www.sdss.org/.

The SDSS is managed by the Astrophysical Research Consortium for the Participating Institutions. The Participating Institutions are the American Museum of Natural History, Astrophysical Institute Potsdam, University of Basel, University of Cambridge, Case Western Reserve University, University of Chicago, Drexel University, Fermilab, the Institute for Advanced Study, the Japan Participation Group, Johns Hopkins University, the Joint Institute for Nuclear Astrophysics, the Kavli Institute for Particle Astrophysics and Cosmology, the Korean Scientist Group, the Chinese Academy of Sciences (LAMOST), Los Alamos National Laboratory, the Max-Planck-Institute for Astronomy (MPIA), the Max-Planck-Institute for Astrophysics (MPA), New Mexico State University, Ohio State University, University of Pittsburgh, University of Portsmouth, Princeton University, the United States Naval Observatory, and the University of Washington.

This research has made use of the NASA/IPAC Extragalactic Database, which is operated by the Jet Propulsion Laboratory, California Institute of Technology, under contract to NASA and of the GOLDMine database \citep{goldmine}.

\onecolumn

\begin{table}
\caption {GALEX observations of the Coma cluster.}
\[
\label{fields}
\begin{array}{ccccc}
\hline
\noalign{\smallskip}
Field &     R.A.   &   Dec.   &  FUV~int.~time   & NUV~int.~time \\
           &    (J.2000) & (J.2000) & (sec)            &  (sec)     \\
\noalign{\smallskip}
\hline
\noalign{\smallskip}									       
NGA\_DDO154 & 12:54:13.0 &  26:55:31  &  1451. & 1451.  \\
Coma\_MOS03 & 12:55:10.0 &  28:24:00  &  1689. & 2547.  \\
Coma\_MOS04 & 12:57:12.0 &  26:42:00  &  1692. & 2630.  \\
Coma\_MOS05 & 12:57:12.0 &  29:00:36  &  1686. & 2606   \\
Coma\_SPECA & 12:57:36.0 &  27:27:00  &  1350  & 1350   \\
Coma\_MOS06 & 13:00:00.0 &  29:00:36  &  1693. & 3524.  \\
Coma\_MOS08 & 13:01:58.0 &  28:36:00  &  1693. & 2598.  \\
Coma\_MOS09 & 13:02:46.0 &  27:48:00  &  1696. & 1696   \\
Coma\_MOS10 & 13:04:00.0 &  29:36:54  &  1694. & 2595.  \\
Coma\_MOS11 & 13:05:19.2 &  28:18:00  &  1699. & 1699.  \\
Coma\_MOS12 & 13:06:14.4 &  27:18:00  &  1701. & 2604.  \\
\noalign{\smallskip}
\hline
\end{array}
\]
\end{table}

\begin{table} 
\caption {Best Fitting Parameters for the NUV and FUV LFs.}
\label{bestfit}
\[
\begin{array}{cccc}
\hline
\noalign{\smallskip} 
Band & Sample & \multicolumn{2}{c}{Schechter~Parameters}\\
     &        &     M^{*}    &   \alpha    \\
\noalign{\smallskip}
\hline
\noalign{\smallskip}
NUV  & Coma^{1} &    -18.50 \pm 0.50       &     -1.77^{+0.16}_{-0.13}    \\
NUV  & Abell1367^{2}   & -19.77\pm0.42 & -1.64\pm0.21\\
NUV  & Field^{3}       & -18.23\pm0.11 & -1.16\pm0.07\\
\hline
FUV  & Coma^{1} &    -18.20 \pm 0.80   &    -1.61^{+0.19}_{-0.25}      \\
FUV  & Abell1367^{2}   & -19.86\pm0.50 & -1.56\pm0.19	\\
FUV  & Field^{3}     & -18.04\pm0.11 &-1.22\pm0.07\\
\noalign{\smallskip}
\hline
\end{array}
\]
1. This work.\\
2. Cortese et al. (2005)\\
3. Wyder et al. (2005)
\end{table}




\end{document}